\documentclass[aps,prl,twocolumn,superscriptaddress,
,floatfix]{revtex4}
\usepackage{graphicx}
\usepackage{amsmath,amssymb}
\usepackage{color}

\begin{document}

\title{High extraction efficiency source of photon pairs based on a quantum dot embedded in a broadband micropillar cavity}

\author{Laia Gin\'{e}s}
\affiliation{Department of Physics, Stockholm University, 10691 Stockholm, Sweden}

\author{Magdalena Moczała-Dusanowska}
\affiliation{Technische Physik, Physikalisches Institut and W\"urzburg-Dresden Cluster of Excellence ct.qmat, Universit\"at W\"urzburg, Am Hubland, D-97074 W\"urzburg, Germany}

\author{David Dlaka}
\affiliation{Quantum Engineering Technology Labs, H. H. Wills Physics Laboratory and
Department of Electrical and Electronic Engineering, University of Bristol, BS8 1FD, UK}

\author{Radim Hošák}
\affiliation{Department of Optics, Palack\' y University, 17. listopadu 12, 77146 Olomouc, Czech Republic}

\author{Junior R. Gonzales-Ureta}
\affiliation{Department of Physics, Stockholm University, 10691 Stockholm, Sweden}

\author{Miroslav Je\v zek}
\affiliation{Department of Optics, Palack\' y University, 17. listopadu 12, 77146 Olomouc, Czech Republic}

\author{Edmund Harbord}
\affiliation{Quantum Engineering Technology Labs, H. H. Wills Physics Laboratory and
Department of Electrical and Electronic Engineering, University of Bristol, BS8 1FD, UK}

\author{Ruth Oulton}
\affiliation{Quantum Engineering Technology Labs, H. H. Wills Physics Laboratory and
Department of Electrical and Electronic Engineering, University of Bristol, BS8 1FD, UK}

\author{Sven H\"{o}fling}
\affiliation{Technische Physik, Physikalisches Institut and W\"urzburg-Dresden Cluster of Excellence ct.qmat, Universit\"at W\"urzburg, Am Hubland, D-97074 W\"urzburg, Germany}

\author{Andrew B. Young}
\affiliation{Quantum Engineering Technology Labs, H. H. Wills Physics Laboratory and
Department of Electrical and Electronic Engineering, University of Bristol, BS8 1FD, UK}

\author{Christian Schneider}
\affiliation{Institut of Physics, University of Oldenburg, D-26129 Oldenburg, Germany}

\author{Ana Predojevi\'{c}}
\email{ana.predojevic@fysik.su.se}
\affiliation{Department of Physics, Stockholm University, 10691 Stockholm, Sweden}

\begin{abstract}
The generation of photon pairs in single quantum dots is based on a process that is, in its nature, deterministic. However, an efficient extraction of these photon pairs from a high-index semiconductor host material requires engineering of the photonic environment. We report on a micropillar-based device featuring an extraction efficiency of 69.4(10)$\%$ that is achieved by harnessing a broadband operation suitable for extraction of photon pairs emitted from a single quantum dot.
Opposing the approaches that rely solely on Purcell enhancement to realize the enhancement of the extraction efficiency, our solution exploits a suppression of the emission into the modes other than the cavity mode. Our technological implementation requires modest fabrication effort enabling higher device yields that can be scaled up to meet the growing needs of quantum technologies. Furthermore, the design of the device can be further optimized to allow for an extraction efficiency of 85$\%$.
\end{abstract}


\maketitle

Quantum dots have shown outstanding potential as emitters of quantum light with promising applications in quantum communication \cite{Pan_internet, Heindel, Greve, Gao, Rare} and quantum computation \cite{Pan_boson}, the latter being a possible path to demonstrate quantum advantage. They can be excited resonantly and coherently to generate on-demand single photons \cite{Pan, Somaschi, Tomm} and photon pairs \cite{Boyle, Flissikowski, Stufler, twophoton}. The advances in design of devices that exploit quantum cavity electrodynamics have allowed for high extraction efficiency, rate, and indistinguishablility of single photons \cite{Pan, Somaschi, Tomm} and therefore, further improved the performance and application range of quantum dot- based quantum light sources. At the same time, it was shown that the photon pairs emitted as biexciton-exciton cascade decay exhibit entanglement in polarization \cite{Akopian}, time bin \cite{time-bin}, and hyperentanglement \cite{hyper}. However, the ultimate performance capability of quantum dot-based sources is yet to be addressed: we require an affordable and high-yield solution that allows for photon pairs emitted by a quantum dot to be extracted with high efficiency. 

The success of quantum dots as single photon sources roots in an intrinsically large dipole moment of the emitter combined with engineering of the photonic environment, the latter being a powerful tool to modify the spontaneous emission processes. Embedding quantum dots in photonic structures such as micropillar cavities \cite{santori, Somaschi, Ding, Unsleber}  was identified as a very promising solution: it maximizes the achievable photon rate, ultimately leading to a deterministic single-photon source performance. In addition to micropillars, a device based on a tuneable open cavity was recently introduced \cite{Tomm}. These approaches utilize moderate quality factor (Q factor of 1000 - 10000), which in return yields cavities with narrow linewidth (typically less than a nanometer), which are not suitable for extraction of photon pairs emitted as a biexciton-exciton cascade. However, moderate Q factor cavities enable strong Purcell enhancement of the emission via the cavity mode into free space, and for certain implementations the magnitude of this enhancement (Purcell factor) as high as 7 \cite{He2017,Somaschi} has been shown. The strong Purcell enhancement is the essential design feature of this approach: increasing the emission into the cavity mode facilitates the high extraction of single photons from the device.

It was widely believed, that a reduction of the cavity Q factor (e.g. the increase of the cavity linewidth) would inevitably deteriorate the device performance via decrease of the Purcell factor. Following this reasoning, an implementation of a micropillar cavity structure that is able to efficiently extract photon pairs generated in a exciton-biexciton cascade would be a challenge, as this task requires a cavity with a linewidth that exceeds the biexciton binding energy (commonly within 1-2\,nm range). Consequently, the most successful implementations of pair sources with engineered extraction efficiency exploited photonic waveguide structures that feature very small device diameter \cite{Claudon}, or coupled micropillar molecules \cite{Dousse} that require custom design to match the emission of the emitter. 
Most recently, hybrid circular Bragg grating cavities \cite{Kartik_BE} were introduced as a promising alternative as they can enhance the emission over a spectral range exceeding the biexciton binding energy and they also provide a well-directed far field that allows for efficient photon collection \cite{Wang, Liu}. Nonetheless, the realization of such a device relies on nanoscale lithography in conjunction with hybrid wafer bonding, which is a demanding fabrication process. In addition to this, the truly successful implementation of a circular Bragg grating structure requires strong concentration of the optical near field at the emitter position. Achieving this calls for an utmost accurate emitter-cavity alignment, which often results in a very low yield of functioning devices.  

\begin{figure}[!ht]
\includegraphics[width=0.95\linewidth]{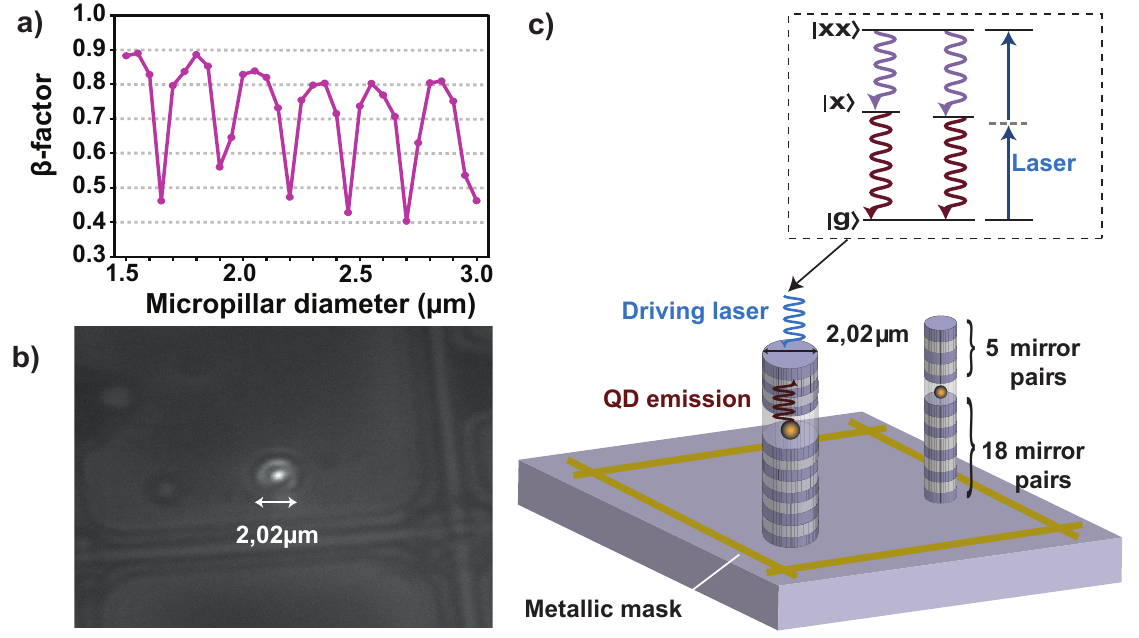}
\caption{ (a) The $\beta$ factor (fraction of emission via the cavity mode vs total emission) shows oscillatory behaviour as a function of the micropillar diameter. The presented analysis was performed for a micropillar with cavity resonance at 910\,nm and a Q factor of 280 \cite{Supplemental}. (b) Image of a micropillar device. The pillar diameter is $2.02\,\mu m$. The bright spot in the center arises from the quantum dot photoluminescence. (c) Schematic of the device and the excitation method. The micropillars are fabricated with a deterministic placement over the site of formation of the quantum dot, hence, the quantum dot is positioned in the center of the micropillar. The top 5 and bottom 18 pairs of $\lambda/4$-thick AlAs/GaAs distributed Bragg reflector mirrors form a cavity with low Q factor. The quantum dot is excited co-linearly with the axis of the micropillar. The inset represents the two-photon resonant excitation scheme \cite{twophoton}. A laser coherently couples the ground, $|g\rangle$, to the biexciton state, $|xx\rangle$, through two-photon resonant excitation. The excitation process is carried out via a virtual state (dashed line). The quantum dot system decays to the ground state via exciton state, $|x\rangle$, emitting the biexciton-exciton cascade.}
\label{pillar_2p}
\end{figure}

In this work, we employ a significantly simpler approach. We utilize a micropillar cavity with a low Q factor and we demonstrate that such a structure features extremely high extraction efficiency. While we show a reduction of the photon emission lifetime (Purcell enhancement) for the cavity embedded emitters, this modest Purcell enhancement taken on its own is not sufficient to explain the extraction efficiency of the device. The phenomenon we observe derives from a combined effect of the Purcell enhancement and a simultaneous suppression of emission into the non-cavity modes. 
The suppression is a result of destructive interference induced for certain ratios of the emitter wavelength and the micropillar lateral size. The existence of this effect was theoretically predicted in \cite{Niels1} for the case of a moderate Q factor micropillar. Our simulations \cite{Supplemental} show that the periodic efficiency modulation also occurs in micropillar cavities featuring low Q factor and that the probability of emission via the cavity mode, commonly characterized by $\beta$ factor \cite{Hughes, b_Niels}, can exhibit very high values (shown in Fig.~\ref{pillar_2p}a). Furthermore, we experimentally demonstrate that our design maintains high performance despite of low Q factor, and that in our device it is possible to achieve simultaneously the high efficiency of pair generation, emission enhancement, and high extraction efficiency of photon pairs. The collection efficiency that we experimentally achieve (69.4(10)$\%$ at the cavity resonance) is extraordinary for a broadband cavity and it indicates that our device design can be used as an example for development of future sources of entangled and correlated photon pairs.

Our device consists of a quantum dot positioned in the center of a micropillar cavity (Fig.~\ref{pillar_2p}b and Fig.~\ref{pillar_2p}c). The relative position of the micropillar with respect to the quantum dot was defined via photoluminescence imaging method \cite{He2017}. To achieve a linewidth of the cavity that is sufficiently large to permit the extraction of both photons (biexciton and exciton) the device was designed to have a Q factor in the range 200-300, which yields a cavity bandwidth of $\sim$5\,nm. The cavity features an asymmetric design (Fig.~\ref{pillar_2p}c), in order to direct the emitted photons towards the top. The details of the growth, positioning, and processing are given in \cite{Supplemental}.

 Opposed to the alternative strategies for a realization of high extraction efficiency pair source, our process is considerably easier to implement. 
 Since the cavity top and bottom mirror reflectivity are very different, the intensity of the emitter luminescence during the imaging is very high. Therefore, it provides an excellent signal to noise ratio, which enables a very accurate spatial positioning ($<$30\,nm). The large bandwidth of the cavity, in turn, alleviates the need for spectral matching with accuracy in range of 10-100 µeV, which is commonly required for devices intended for extraction of single photons. Additionally, the birefringence in a low Q cavity is very small, which is important for preserving polarisation correlations in the biexciton-exciton cascade, required for obtaining a high degree of photon entanglement and high indistinguishability.  

The sample was kept for measurements at 4K, and the quantum dots were excited via two-photon resonant excitation of the biexciton \cite{twophoton}, schematically depicted in the inset of Fig.~\ref{pillar_2p}c. The excitation pulses were derived from a pulsed 80MHz repetition rate Ti:Sapphire laser. The excitation pulse-length was tailored using a pulse-stretcher assembled in a 4f configuration \cite{twophoton}. The laser scattering was filtered out by means of spectral and polarization rejection and the exciton and biexciton emission were separated ahead of single-mode fiber coupling.

\begin{figure}[!ht]
\includegraphics[width=0.95\linewidth]{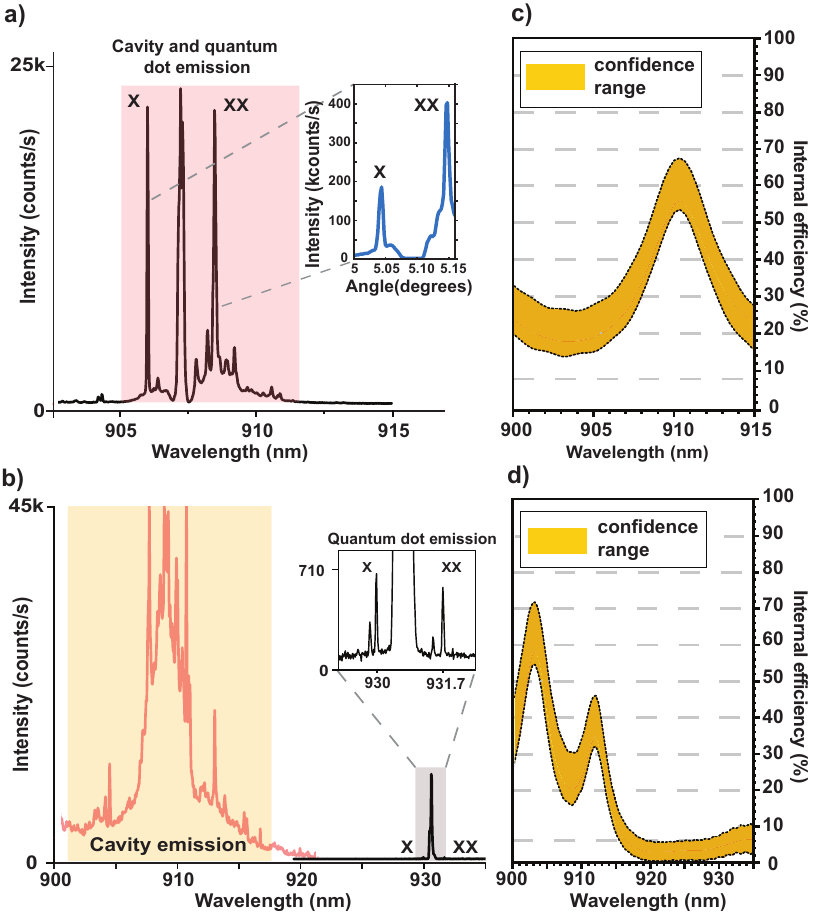}
\caption{(a) Emission spectrum of a quantum dot under two-photon resonant excitation. Both the biexciton and the exciton emission lie within the linewidth of the cavity. The biexciton emission at $908.5\,nm$ is overlapped with the cavity resonance. The micropillar device employed is shown in the  Fig.~\ref{pillar_2p}b, featuring a diameter of $2.02\,\mu m$. The inset shows a scan of the same spectrum obtained using a single mode fibre coupled avalanche photodiode as a function of the angle of rotation of a diffraction grating. The single mode fibre acts as spatial filter favoring the cavity mode emission. (b) Photoluminescence of the cavity with a diameter of $2.75\,\mu m$ achieved in above-band excitation and the emission spectrum of a quantum dot far detuned from this cavity obtained in two-photon resonant excitation. The inset shows a close-up of the quantum dot emission. The laser scattering is visible between the exciton (x) and biexciton (xx) emission lines. (c) Internal efficiency as a function of wavelength for micropillar with diameter $2.02\,\mu m$. The confidence range is derived taking into account the simulation errors and fabrication accuracy. (d) Internal efficiency as a function of wavelength for micropillar with diameter $2.75\,\mu m$. The shorter wavelength centered at $905\,nm$ is a higher order mode. In the Fig.~\ref{spectra}b this mode is filtered by the optics.}
\label{spectra}
\end{figure}

The deterministic positioning of micropillars, which was introduced above, has allowed us to identify devices that feature an evident spectral fingerprint of the biexciton-exciton emission cascade. The seven structures, which were tested in the course of this study, showed a quantum dot emission that was lying within the linewidth of the cavity mode. In turn, we selected two reference micropillars, where the quantum dot emission was far detuned from the cavity resonance (low energy detuning with respect to the cavity mode of 0.0236\,eV (15\,nm) and 0.0322\,eV (22.59\,nm), respectively. Figure \ref{spectra}a shows the emission spectrum of a quantum dot where the biexciton line is resonant to the cavity. The Fig.~\ref{spectra}b shows the emission of a quantum dot far detuned from the cavity resonance. Comparing the spectra shown in Fig.~\ref{spectra}a and Fig.~\ref{spectra}b one can observe a strong difference in the intensity of emission lines. This result is in agreement with the simulation \cite{Supplemental}, shown in Fig.~\ref{spectra}c and Fig.~\ref{spectra}d, respectively. The Fig.~\ref{spectra}c and Fig.~\ref{spectra}d show the spectral profiles of the internal efficiency, the portion of the quantum dot emission that is, via the cavity mode, emitted through the top of the device. While the internal efficiency is very high at the cavity resonance, it is low far away from the cavity resonance.

\begin{figure}[b]
\includegraphics[width=0.93\linewidth]{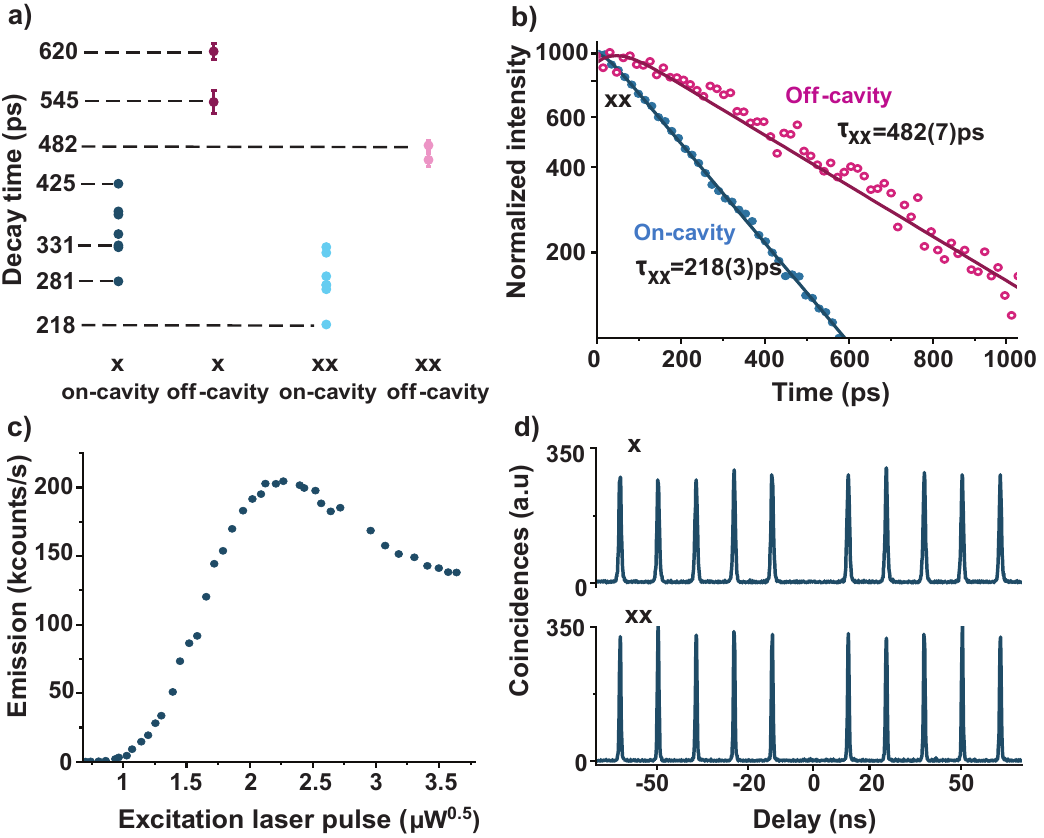}
\caption{(a) Summary of lifetime values obtained for all the measured devices (b) Lifetime of biexciton photon for an on-cavity and an off-cavity emitter (c)  Rabi oscillations of the biexciton. (d) The autocorrelation measurements of the exciton and biexciton photons yielding $g^{(2)}(0)_{x}$=0.016(3) and $g^{(2)}(0)_{xx}$=0.009(2).}
\label{lifetime}
\end{figure}

To further estimate the effect of the cavity on the emitters we carried out lifetime measurements. To obtain the values of lifetimes we first determined the system response function by detecting the attenuated laser signal derived from a pulsed laser. The system response includes the single photon detector response, the laser pulse duration and the acquisition system response. It was estimated fitting a Gaussian function to the recorded signal, yielding a value of $\tau_{S}$=78(2)\,ps, full width at half maximum. The biexciton lifetime was obtained fitting the convoluted function of the system response function and a single-exponential function to the experimental data. The acquired biexciton lifetime values for cavity resonant quantum dots were in a range between $\tau_{xx}$=218(3)-331(3)\,ps. On the other hand, to determine the lifetime of exciton photons, two single-exponential decay functions were convoluted, and the previously estimated biexciton lifetime was used as the fast component parameter. The exciton lifetimes in the cavity were in a range between $\tau_{x}$=281(4)-425(4)\,ps. For the quantum dots that were far-detuned from the cavity, the lifetime values were also determined following the same fitting method, yielding lifetimes values of $\tau_{x}$=620(12)\,ps and $\tau_{x}$=545(17)\,ps for the excitons and $\tau_{xx}$=482(7)\,ps and $\tau_{xx}$=460(10)\,ps for the biexcitons, respectively. In Fig.~\ref{lifetime}a we show the lifetime measurements for all nine devices that we investigated. The measurements employing the seven devices where the quantum dot emission was within the linewidth of the cavity are denoted as on-cavity, while the remaining two devices that were far detuned from the cavity are denoted as off-cavity. 

The micropillar device that we investigated further had the diameter of $2.02\,\mu m$. Our theoretical assessment (shown in Fig.~\ref{pillar_2p}a) predicts that a micropillar with such a diameter should feature the $\beta$ factor higher than 0.8, which motivated out choice of the device (also shown in Fig.~\ref{spectra}a). The lifetime of the cavity-resonant biexciton is shown in Fig.~\ref{lifetime}b. The same figure shows, for comparison, the lifetime of the biexciton emitted by the quantum dot far detuned from the cavity resonance (spectrum shown in Fig.~\ref{spectra}b). The coherent nature of the excitation was confirmed by Rabi oscillations shown in Fig.~\ref{lifetime}c. In subsequent measurements we probed the statistics and indistingishablity of the emitted photons. A very strong sub-Poissonian and anti-bunched statistics of the emitted state was demonstrated by means of autocorrelation measurements under $\pi$-pulse excitation. Here, by $\pi-$pulse we denote the pulse area of the excitation pulse that excites the quantum dot with the highest probability attainable (Fig.~\ref{lifetime}c). Low multiphoton contribution (Fig.~\ref{lifetime}d) was observed at zero delay, with $g^{(2)}(0)_{x}$=0.016(3) and $g^{(2)}(0)_{xx}$=0.009(2) for exciton and biexciton, respectively. To test the indistinguishibility of the consecutively emitted photons, Hong-Ou-Mandel (HOM) interference experiments were performed. From the experimental data we obtained a $g^{(2)}_{xx HOM}$= 0.280(10) which brings on two photon interference visibility of 0.47(3) for biexciton photons, a value not far off from what is predicted by theory \cite{simon, huber_master}. 

The outstanding potential of our device was unequivocally confirmed by measurement of the photon collection efficiency. While applying the laser with the repetition rate of 80MHz and a pulse area that maximizes emission probability ($\pi$-pulse), we observed count rates on avalanche photodiodes (APD) of 401(1) kcounts/s and 198(1) kcounts/s for the biexciton and exciton photons, respectively. The biexciton binding energy separates the biexciton and exciton emission channels by 2.4\,nm. The efficiency of the APDs we used was $\eta_{APD}$=0.4 while the single mode fibre coupling efficiency was measured to be $\eta_{fibre}$=0.291(10) for both biexciton and exciton photons. The setup efficiency (overall optical path transmission) was measured to be $\eta_{optics}$=0.062(10). The blinking of the quantum dot was not observed. Using these figures, we obtained a collection efficiency of 69.4(10)$\%$ and 35.1(10)$\%$ for biexciton and exciton photons collected into the first lens above the sample (NA=0.68), respectively. Our simulation of the emission far field indicates that the device numerical aperture is of approximately 0.4. This translates the collection efficiency acquired using a lens with NA=0.68 into a direct measurement of the internal efficiency. 

\begin{figure}[t]
\includegraphics[width=0.92\linewidth]{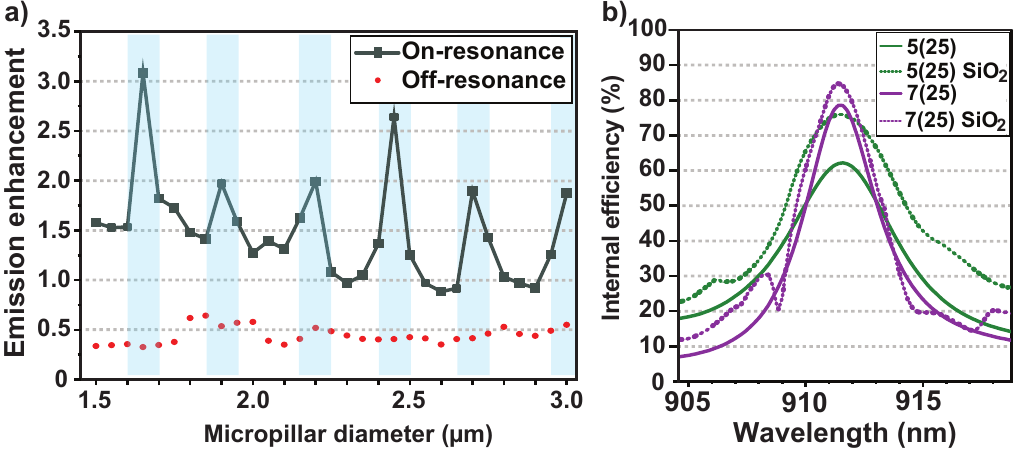}
\caption{(a) Emission enhancement on- and off-resonance. The blue shaded area of the strong enhancement corresponds to the micropillar diameters where we expect the reduction of the $\beta$ factor (Fig.~\ref{pillar_2p}a). Such a behavior roots in the presence of the side-modes that, whilst representing an enhancement of the emission rate, actually act to deviate the emission out of the sides of the device. The detail description on how this graph is achieved is given in the \cite{Supplemental}. The green points pertaining to the while area involve the emission via the cavity mode only, and as such are on-resonance values of the Purcell factor (b) Further improvement of the internal efficiency simulated for a micropillar with a diameter $1.8\,\mu m$. The four shown results feature design improvements including increased number of bottom DBRs from 18 to 25, a SiO$_{2}$ substrate, and a minor reduction of the cavity Q factor achieved by increasing the number of top DBR's from 5 to 7.}
\label{model}
\end{figure}

The high internal efficiency presented here (69.4(10)$\%$ at the resonance) was not achieved by exploiting exclusively the Purcell enhancement of the cavity resonant emitters. The underlying process is more complex, it joins the effects of Purcell enhancement and the suppression of emission into the modes other than the cavity mode. We still observe a lifetime reduction of emitters resonant to the cavity compared to the lifetime of the emitters that are far detuned from the cavity. The theoretically expected rate enhancement for a micropillar with a diameter of $2\,\mu m$ (Fig.~\ref{model}a) is $\approx$1.25, which on its own would yield an internal efficiency of 55$\%$ at the cavity resonance. However, our theoretical simulation indicates that the rate suppression at the wavelengths that lay outside the cavity linewidth also plays a fundamental role (red points in Fig.~\ref{model}a). Namely, on average the Purcell factor out of the cavity mode is ~0.5. This leads to a roughly a factor 2 difference between the on vs off-cavity emission rates, which matches both the lifetime and efficiency data we measured.

We have demonstrated a device achievable by a simple and reproducible fabrication method. The device features a large linewidth cavity suitable for collection of photon pairs emitted via biexciton-exciton cascade. The adequate choice of the micropillar diameter enables high collection efficiency unprecedented for a broadband micropillar-based device. Furthermore, this number can be improved by means of an optimized design, that maximizes the internal efficiency by reducing the portion of the cavity emission directed downwards into the GaAs substrate. We investigated how the internal efficiency can be improved for a $1.8\,\mu m$ diameter pillar, as such diameter yields a theoretical value of $\beta$ factor of 0.9. We studied four different cases shown in Fig.~\ref{model}b. The internal efficiency can be significantly improved by increasing the number of bottom DBRs and use of a low refractive index substrate.
Last but not the least, our device can be employed to generate entangled photon pairs. The micropillar structure can be subjected to strain tuning \cite{tuning1, tuning} with a goal to generate polarization entangled photon pairs.

\begin{acknowledgments}
We acknowledge funding by the DFG within the projects SCHN1376-5.1 and PR1749/1-1. We acknowledge financial support by the State of Bavaria. A.P. would like to acknowledge Swedish Research Council and Carl Tryggers Stiftelse. J.R.G.U., R.H., and M.J. are supported by project HYPER-U-P-S.  Project  HYPER-U-P-S  has  received  funding  from  the QuantERA ERA-NET  Cofund in  Quantum  Technologies  implemented  within  the  European  Union’s  Horizon  2020  Programme. L. G. was supported by the
Knut \& Alice Wallenberg Foundation (through the Wallenberg Centre for Quantum Technology (WACQT)). A.Y. acknowledges support from EP/N003381/1 One-dimensional quantum emitters and photons for quantum technologies: 1D QED. M.J. would like to acknowledges funding by the MEYS within the project 8C18002 (HYPER-U-P-S) and the Czech Science Foundation within the project 21-18545S. 
\end{acknowledgments}

\end{document}